\documentclass[12pt,preprint]{aastex}

\begin{document}

\title{A NEW DISTANCE ESTIMATOR AND
SPATIAL DISTRIBUTION OF  GRBS OBSERVED BY BATSE}

\author{Heon-Young Chang$^1$, Suk-Jin Yoon$^2$, and Chul-Sung Choi$^3$}

\affil{$^1$
Korea Institute for Advanced Study\\
207-43 Cheongryangri-dong Dongdaemun-gu, Seoul 130-012, Korea,\\
$^2$
Center for Space Astrophysics and Department of Astronomy,
           Yonsei University\\
 134 Shinchon-dong Seodaemun-gu, Seoul 120-749, Korea\\
$^3$
Korea Astronomy Observatory \\
36-1 Hwaam-dong, Yusong-gu, Taejon 305-348, Korea}
\email{hyc@ns.kias.re.kr, sjyoon@csa.yonsei.ac.kr, cschoi@kao.re.kr}

\begin{abstract}
We propose a redshift estimator for the long  ($T_{90} > 20$) 
gamma-ray bursts (GRBs) observed by the BATSE. 
It is based on an  empirical relation between
the redshift and the power-law index of power density spectra (PDSs)
of the observed GRBs. This relation is constructed by using the fact that
the power-law index is dependent upon
a characteristic timescale of GRB light curves which are 
inevitably stretched by cosmological time dilation. 
We construct the empirical relation using 
both individual PDSs and  averaged PDSs.
An error estimates of $z$ are 1.09 and 1.11 for the empirical relation
by individual PDS fits, 
1.72 and 1.56 by averaged PDS fits, 
for the least squares fit and the maximum likelihood fit, respectively.
We attempt to determine the spatial
distribution of the GRBs observed by the BATSE as a function of 
redshifts on the basis of the resulting redshift estimator. 
We find that the obtained spatial distribution of the 
observed GRBs seems consistent with that of the GRBs whose redshifts
are reported, even though the estimated errors are not very accurate.
The GRBs observed by the BATSE seem distributed within $z \sim 5-6$. 
This result has implications on theoretical calculations
of stellar formations at high redshifts and beaming geometry via 
a statistical study of the observed GRBs involving 
beaming-induced luminosity functions.
We discuss such implications, and possible 
uncertainties of the suggested method. 
\end{abstract}

\section{Introduction}

Since gamma-ray bursts (GRBs) were first discovered in late 60's
(Klebesadel et al. 1973), 
thousands of GRBs have been detected up to date. 
The discovery of afterglows 
in other spectral bands and host galaxies enabled us to
measure redshifts of about twenty GRBs
(see, e.g., ${\rm http://www.aip.de/\sim jcg/grbgen.html}$),
establishing the fact that GRBs 
are indeed cosmological 
(Mao and Paczy${\rm \acute{n}}$ski 1992; Meegan et al. 1992;
Piran 1992; Metzger et al. 1997).
Nonetheless, the redshifts are still unknown
for most of the detected GRBs.  Unless we locate
a burst on the sky by immediate follow-up observations, the distance
of the burst is apt to remain unrevealed forever. 

Concerning physical models of GRBs,
their distance scales are related to key issues, 
such as, energetics, burst rates as a function of redshifts.
That is, estimating the distance puts direct constraints
on the theories of the observed GRBs.
Besides this
the redshift distribution of GRBs should track the cosmic star 
formation rate of massive stars,
if  GRBs are indeed related to the collapse of massive stars
(Woosley 1993; Paczy${\rm \acute{n}}$ski 1998; MacFadyen and Woosley 1999).
Therefore, once its association has been proven, one expects 
the observed GRBs are the most powerful probe of the high redshift universe
(Wijers et al. 1998; Blain and Natarajan 2000; Lamb and Reichart 2000).
In fact, the GRB formation rate and the star formation rate (SFR) 
have similar slopes at low redshift, 
implying that GRBs can be used indeed
as a probe of the cosmic star formation rate at high redshift. 
Observations of faint galaxies have been used to estimate the history of 
star formation activity (Madau et al. 1996; Rowan-Robinson 1999; 
Steidel et al. 1999). However, there are considerable difficulties and
uncertainties in the corrections that should be applied due to
extinction and obscuration. An independent determination and 
test of the relative amount of obscured and unobscured star formation 
activity would be extremely valuable.

At present, there are too few redshift
measurements with which to produce the global GRB formation rate. 
This fact is indeed hard to avoid unless observers set up networks
of efficient telescopes in order for an immediate follow-up observations. 
Recently, however, there are pilot studies 
to overcome the technical ability mentioned above 
(e.g., Reichart et al. 2001), though
there have been several attempts to quantify pulse shapes of GRBs 
and interpret  results in terms of GRB physics 
(Fenimore et al. 1996; Norris et al. 1996; In'T Zand and Fenimore 1996;
Kobayashi et al. 1997; Daigne and Mochkovitch 1998; Fenimore 1999;
 Panaitescu et al. 1999). 
Several authors (Stern et al. 1999; Fenimore and Ramires-Ruiz 2000;
Reichart et al. 2001) 
began to observe strong correlations between temporal
properties of the observed GRBs and their brightness, which may
have some implications that the measured 
spikiness can be used to obtain 
distances much like a Cepheid-like distance estimator.
Norris et al. (2000) also  showed the spectral lag/luminosity relationship for 
six bursts with known redshifts can be appreciated.
Currently, the luminosity estimator yields  best-estimate 
luminosity distances that are accurate to a factor 
of $\approx 2$ (see Reichart et al. 2001).

Along the line of efforts of such kinds
we propose a new method based on an  empirical relation motivated
by the work of Chang (2001). 
It is well known that power density spectra
(PDSs) of long GRBs show a power-law behavior (Beloborodov et al. 1998, 2000). 
Though its underlying physical mechanism is not obvious 
(Panaitescu et al. 1999; Chang and Yi 2000),
the PDS analysis  may provide useful information of physics of GRBs 
(Panaitescu et al. 1999) and the distance information (Chang 2001).
Particularly, Chang (2001) has demonstrated that the power-law 
index of PDSs of the observed GRBs shows a redshift dependence, 
implying a possible relationship between the power-law
index and the redshift of GRBs. 
It can be possibly worked out because of the fact that
burst profiles should be stretched in time due to 
cosmological time dilation by an amount proportional to the redshift, $1+z$.

In \S 2 we begin with a brief summary of the PDS analysis in GRB studies,
and describe the  empirical relation involved in our procedure. 
In \S 3 we present results obtained by
applying our method to the GRBs observed by the BATSE instrument aboard 
the {\it Compton Gamma Ray  Observatory}  (Paciesas et al. 1999),
and discuss what they suggest.
Finally, we conclude by pointing out that the accuracy of our 
redshift estimates is  limited by unknown underlying properties of 
GRBs and what should be further developed in \S 4.

\section{Empirical Relation of Power-law Index and Redshift}

Contrary to the diverse and stochastic behavior in the time domain,
long GRBs show a simple behavior in the frequency domain
(e.g., Beloborodov et al. 1998). The power-law
behavior is seen even in a single burst when it is bright and long.
The power-law PDS provides a new tool for studies of GRBs
themselves. Using the PDS analysis, Panaitescu et al. (1999) analyzed 
the temporal behavior of GRBs in the framework of a relativistic 
internal shock model. They set up 
their internal shock model and attempted to identify the most sensitive model
 parameters to the observed  PDS,
which is defined by the square of
the Fourier transform of the observed light curve.
They concluded that
the wind must be modulated such that collisions at large radii release 
more energy than those at small radii in order to reproduce consistent 
PDSs with the observation. However, it is also noted that
the reported power-law behavior with the index of $-5/3$ and 
the exponential distribution of  the observed PDS
can be reproduced by adjusting the sampling interval in the time domain 
for a given decaying timescale of individual
pulse in a specific form of GRB light curves (Chang and Yi 2000).
Therefore, conclusions on the central engine of GRBs on the basis of
the PDS analysis should be derived with due care. 

Another valuable use of the PDS analysis can be realized bearing in mind
that for a given sampling interval 
the resulting power-law index is dependent upon
the characteristic timescale of the observed light curve. For instance,
consider a GRB light curve as a sum of exponential functions of time, $f(t)$,
as considered in Chang and Yi (2000). Since the
Fourier transform of $f(at)$ is given by $\frac{1}{|a|}F(\nu /a)$, where
$F(\nu)$ is the Fourier transform of $f(t)$, 
$\nu$ being the cyclic frequency and $a$ being
a constant, fitting of the power law function to the PDS 
may result in a different power-law index 
as the constant $a$ varies when the sampling interval is pre-determined. 
In other words, for the observed GRB light curves with the pre-determined
sampling intervals, e.g., 64 ms, the cosmological time dilation stretches
the light curve by an amount of $1+z$ 
and consequently results in changes in the
obtained power-law index. This should be true because cosmological objects 
like GRBs should not only be redshifted in energy but also extended
in time because of the expansion of the universe. 
Chang (2001) demonstrated that a cosmological time dilation effect is 
indeed imprinted in the light curves of the observed GRBs
whose redshifts are known by dividing
the GRBs into near and far groups. The author has showed that the near
GRB group ends up with the smaller 
power-law index than the far one and that
the correction with the $1+z$ factor removes the differences. 
The power-law index
difference in two separate groups is larger than  
that among different energy bands. 

In order to construct the empirical relation between the power-law index
and the redshift of GRBs we have calculated the power-law index of the PDSs 
of 9 GRBs detected by the BATSE with known redshifts. 
We have used light curves of the  GRBs from the 
updated BATSE 64 ms ASCII 
database\footnote{ ${\rm ftp://cossc.gsfc.nasa.gov/pub/data/batse/}$}. From this
archive we select the light curves of the GRBs in channel 2 
whose redshifts are available.
We list up the GRBs used in our analysis with 
their reported redshifts in Table 1.
We calculate 
the Fourier transform of each light curve of the GRBs
and the corresponding PDS.
Before taking the Fourier transform of 
light curves we scale them such that the height of their highest peak has 
unity in the  GRB light curves.
Since the individual PDS of GRBs are stochastic,  
different parts of the PDS appear to follow a slightly different power-law index.
Having calculated the PDS of an individual GRB we obtain the power-law index
of the PDS using the limited part of the PDS, i.e., $-1.6 < \log \nu < 0$.
The lower bound is roughly determined in  that the deviation from the 
power law begins due to the finite length of bursts.
The upper bound is where the Poisson noise 
becomes dominant. 
Poisson noise in the measured count rate affects the PDS at high frequencies 
and has a flat spectrum. 
The Poisson noise level equals the burst total fluence including the 
background in the considered time window. We calculate the individual 
Poisson level for each burst and subtract it from the burst PDS.

The PDSs can be described as a single power law with superimposed 
fluctuations which follow the exponential distribution,
which may require the maximum likelihood method. However, by considering
that we smooth the PDSs on the scale $\Delta \log \nu = 0.5$ before fitting,
the least squares fit can be preferred since the error distribution
may be modified to the normal distribution according to the 
{\it central limit theorem}. We use two different fitting 
routines corresponding to the normal error distribution and the 
exponential error distribution. 
We employ both fitting algorithms to compare.
What is shown in Figure 1 are  empirical relations of the redshift 
and the power-law index obtained by the least squares fit 
and the maximum likelihood method.
Firstly, we attempt to construct the empirical relation of the
redshift and the power-law index without averaging of the PDSs. 
That is, we construct the best
fit using power-law indices of 9 individual data points. 
Results are also shown in Figure 1, where 
the thin solid line results from the least squares fit, the thin dashed
line the maximum likelihood method. 
A  possible alternative way to extract the empirical relation
from the noisy individual PDSs is to take the average PDSs over a sample 
of long GRBs. Then the random fluctuations affecting each individual PDS tend 
to cancel each other and the power-law behavior can be clearly seen.
Because of  the small number of data, we group the PDSs into 4 subgroups
according to the reported redshifts, and average the PDSs before a fitting
process :
GRB 980329 + GRB 971214, GRB 990123 + GRB 990506 + GRB 990510, GRB 970508
+ GRB 980703 + GRB 991216, GRB 980425. 
The thick solid line results from fitting of averaged PDSs by
the least square fit, the thick dashed
line the maximum likelihood method.
We note that individual fitting results in a  less steeper
relation. We also have attempted higher order polynomial fits but it did not
end up with a monotonic relation as one should expect.

The error estimates of $z$, which are defined by a square root of 
the average of squared difference between
the measured redshifts summarized in Table 1 
and the expected redshifts by the fitting, 
are 1.09 and 1.11 for the empirical relation due to individual PDS fits, 
1.72 and 1.56 due to the averaged PDS fits, 
for the least squares fit and the maximum likelihood fit, respectively.

\section{Spatial Distribution of GRBs}

We adopt light curves of the long GRBs from the updated BATSE 64 ms ASCII 
database as in processes above. 
We choose bursts with durations $T_{90} > 20 $ s, where $T_{90}$ is 
the time it takes to accumulate from 5 \% to 90 \% of the total fluence 
of a burst summed over all the four channels. Of those bursts, 
we further select bursts with the
peak count rates satisfying $C_{\rm max}/C_{\rm min}$ 
for the 64 ms trigger timescale is greater than 1.
Applying these criteria, we end up with 388 bursts.

In a similar way, we obtain power-law indices of PDSs of the selected
GRBs and subsequently estimates of their redshifts.
In Figure 2, the redshift distributions of the GRBs obtained by
the relation we have in the previous section are shown. Different line types
indicate same meanings as in Figure 1. Note that the thin dotted histogram 
represents the spatial distributions of the 22  GRBs whose redshifts are
available at the web site, from where the quoted redshifts in Table 1 are taken.
It is interesting to note that the predicted redshift distributions of 
GRBs  that are  derived from fitting of individual power-law index 
without averaging PDSs appear to provide a  better agreement with 
the redshift distribution of the GRBs with redshift-known. 

The obtained GRB redshift distribution is quite suggestive. Firstly,
it is seen in Figure 2 that the long GRBs observed by the BATSE, at least,
are distributed well within $z \sim 5-6$. 
If one accepts an idea that the GRB formation rate should trace
the SFR at high redshifts as at low redshifts, this is in apparent 
contrast to what is derived from some theoretical calculations of star formation.
Theoretical calculations show that the birth rate of Pop III stars produces 
a peak in the SFR in the universe at redshifts $16 \la z \la 20$, 
while the birth rate of Pop II stars produces a much larger and broader 
peak at redshifts  $2 \la z \la 10$ (Ostriker and Gnedin 1996). Secondly,
according to the cumulative redshift distributions derived in terms of 
beaming-induced luminosity functions an extreme shape of a conic 
beam seems likely to be ruled out : although the observed $<V/V_{\rm max}>$
can be satisfied with the theoretical $<V/V_{\rm max}>$,
a broad beam cannot explain the observed redshifts greater than $\sim 2 - 3$ 
(Kim et al. 2001) and a hollow beam expects too many GRBs farther than
$z \sim 5-6$ (Chang and Yi 2001),  if  a SFR-motivated
number density distribution of GRB sources is assumed.

\section{Discussions}

There are problems and limits to determine redshifts accurately
in both obtaining the relation and applying this relation to data. 
First of all, even though this method is in principle to 
work, it is not clear whether we may apply this method over the observed 
GRB light curves obtained by various satellite missions at the same time.  
It is partly because the PDS of each individual burst is composed of 
the power law and superimposed exponentially distributed fluctuations
which make it difficult to recognize the power law in an individual burst,
and  partly because such the empirical relation is susceptible to 
observational conditions, such as, trigger timescale, detection sensitivity.
Availability of more redshifts of GRBs may help to reproduce a better
relation.
With all the efforts in implementing a sophisticated algorithm to accommodate 
the diversity of the light curves, it is essential to understand
a fundamental mechanism of GRBs to derive 
the intrinsic relation of the power-law index and the redshift.
Secondly, We have implicitly assumed that 
all the long GRBs have a more or less
same characteristic timescale and cosmological time dilation alone
affects varying the characteristic timescale.
We need to understand clearly  what and how forms the flat part of PDS.
Thirdly, the effect of redshift tends to flatten PDSs of GRBs
on the contrary to the time dilation effect. 
It reflects a well-known fact that  pulses in 
a single  GRB are more narrow in a higher energy band 
(e.g., Norris et al. 1996). 
These effects combine and produce undesirable results in obtained
power-law index.  We have presumed in this study 
that the time dilation effect on the power-law index is larger than
that of the redshift as observed in Chang (2001)  
and ignored the effect of the redshift.
However, it should be understood how
the power-law index relates with the energy channels
to improve the proposed method accommodating the redshift effect. 

We conclude by pointing out that, even though redshift estimates
are subject to the stochastic nature of the observed PDSs and
accuracy of estimates are limited by unknown properties of the GRBs
the encouraging conclusion of this study is 
that redshifts of the GRBs can be obtained with the GRB light curves, 
whose redshifts otherwise remain unknown forever.

\acknowledgements

SJY is supported by the Creative Research Initiatives Program of
the Korean Ministry of Science and Technology. 
This research has made use of data obtained through 
the High Energy Astrophysics Science Archive 
Research Center Online Service, 
provided by the NASA/Goddard Space Flight Center.

\newpage
\begin{figure}
\plotone{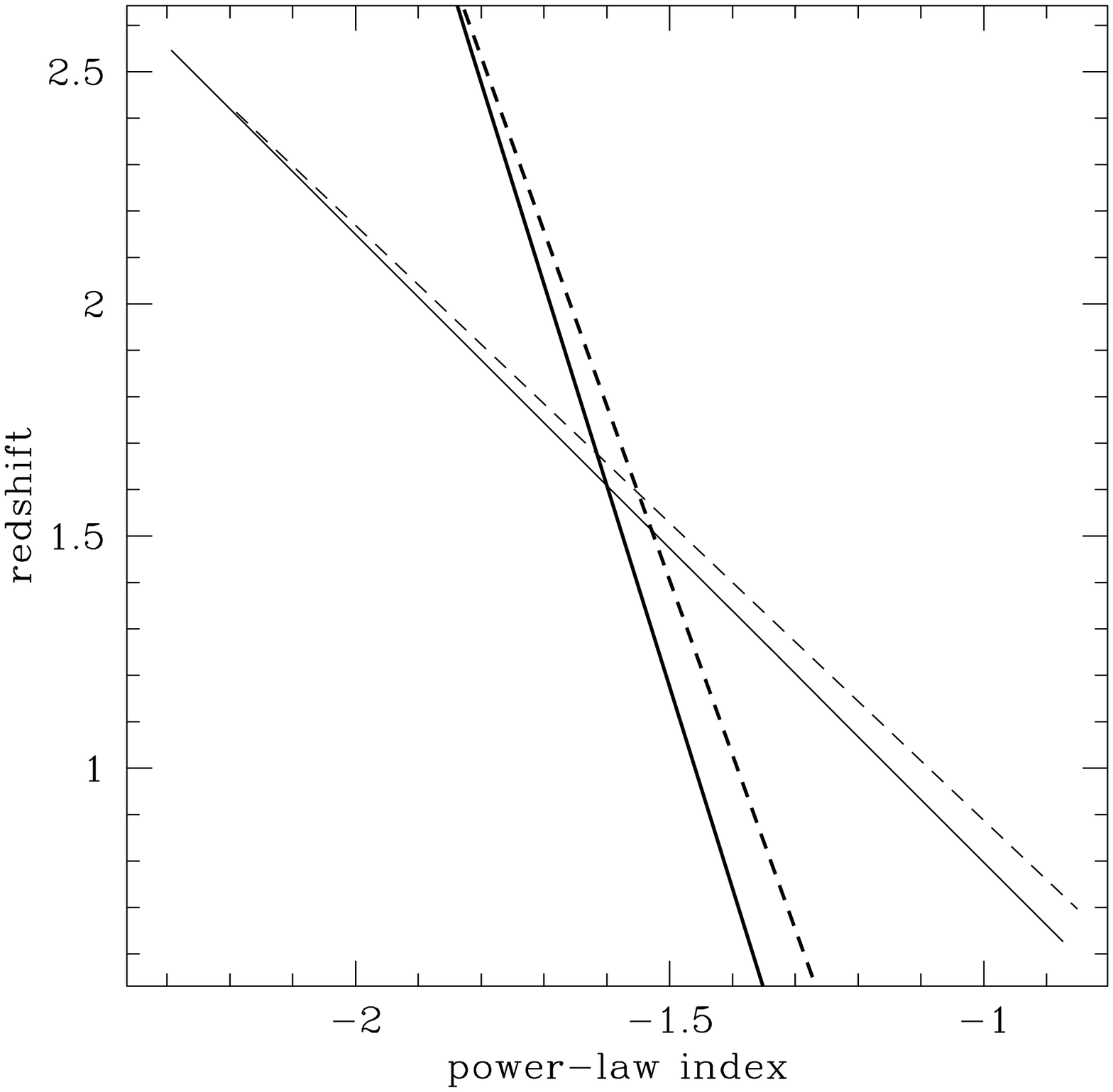}
\caption{
Relation of power-law index and  redshift. 
The thin solid line results from fitting of individual PDSs
by the least squares fit, the thin dashed
line the maximum likelihood method.
For comparison, the relation of 
redshift and  power-law index with averaged PDSs is also shown. 
The thick solid line and the thick dashed line represent 
the least squares fit, the maximum likelihood method, respectively. 
\label{fig1}}
\end{figure}

\begin{figure}
\plotone{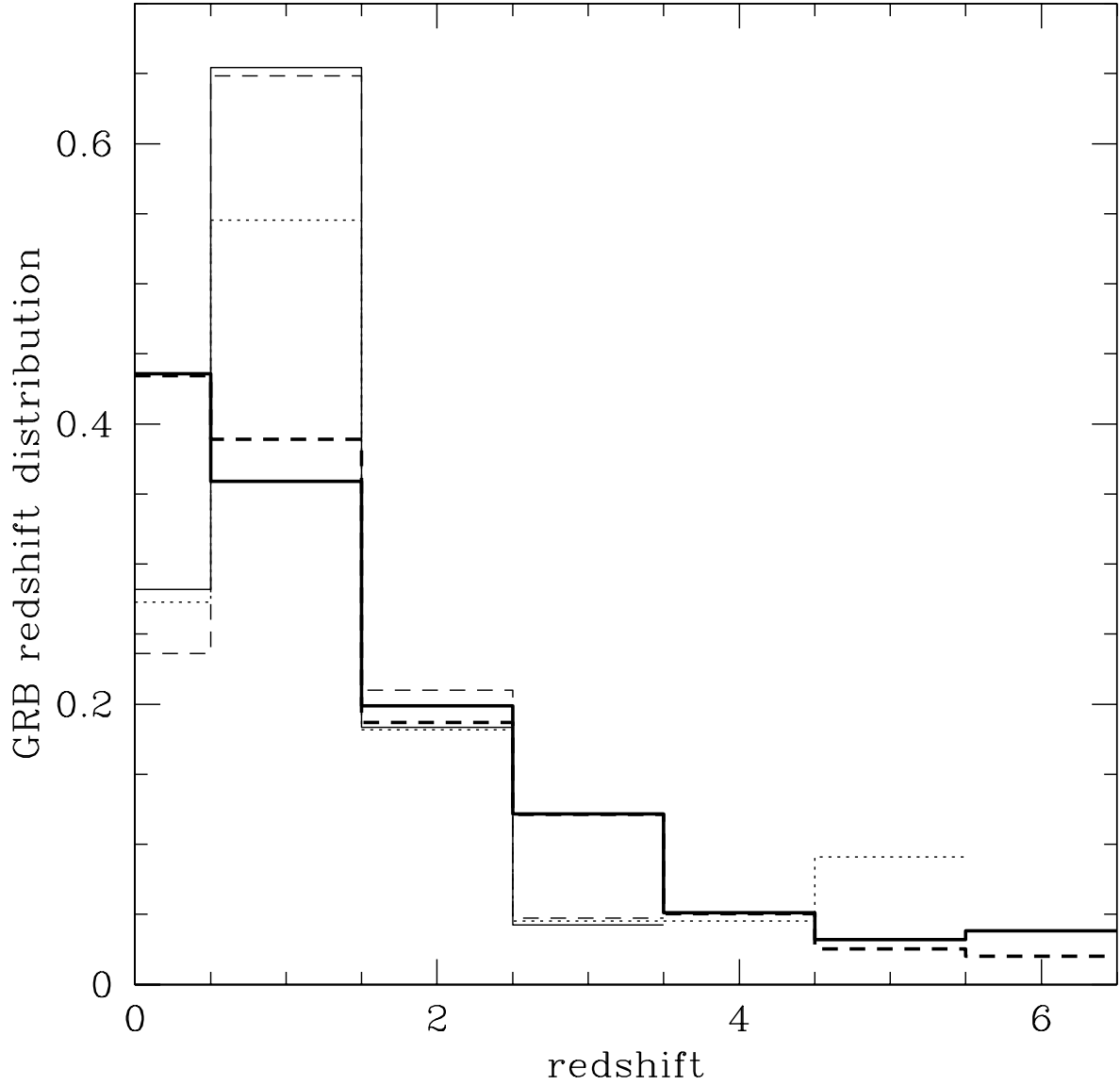}
\caption{
Normalized
spatial distributions of 388 long GRBs observed by the BATSE are compared 
with that of the redshift-known GRBs, indicated by the same line types 
as in Figure 1 and the thin dotted line, respectively.
\label{fig2}}
\end{figure}

\newpage
\begin{table}
\begin{center}
\caption{A list of the GRBs used in the analysis with 
the redshifts and $T_{90}$. 
The redshifts are quoted from a complied table at
${\rm http://www.aip.de/\sim jcg/grbgen.html}$. \label{tbl-1}}
\vspace{2mm}
\begin{tabular}{cccc}
\hline
GRB name&trigger number&redshift&$T_{90}$ (secs) \\ \hline\hline
GRB 991216 &  7906 & 1.02   & 15 \\
GRB 990510 &  7560 & 1.619  & 68 \\
GRB 990506 &  7549 & 1.3    & 130 \\
GRB 990123 &  7343 & 1.60   & 63 \\
GRB 980703 &  6891 & 0.966  & 411 \\
GRB 980425 &  6707 & 0.0085 & 34 \\
GRB 980329 &  6665 & 3.9    & 18 \\
GRB 971214 &  6533 & 3.42   & 31 \\
GRB 970508 &  6225 & 0.835  & 23 \\ \hline
\end{tabular}
\end{center}
\end{table}


\newpage
\begin{thebibliography}{}
\bibitem[]{} 
     Blain, A. W. and Natarajan, P. 2000, MNRAS, 312, L35
\bibitem[]{} 
     Beloborodov, A., Stern, B., and Svensson, R. 1998, ApJ, 508, L25
\bibitem[]{} 
     Beloborodov, A., Stern, B., and Svensson, R. 2000, ApJ, 535, 158
\bibitem[]{} 
     Chang, H.-Y. 2001, ApJ, 557, L85
\bibitem[]{} 
     Chang, H.-Y. and Yi, I. 2001, ApJ, 554, 12
\bibitem[]{} 
     Chang, H.-Y. and Yi, I. 2000, ApJ, 542, L17          
\bibitem[]{} 
     Daigne, F. and Mochkovitch, R. T. 1998, MNRAS, 296, 275
\bibitem[]{} 
     Fenimore, E. 1999, ApJ, 518, 375
\bibitem[]{} 
     Fenimore, E. E. and Ramirez-Ruiz, E. 2000, astro-ph/0004176
\bibitem[]{} 
     Fenimore, E., Madras, C. D., and Nayakshin, S. 1996, ApJ, 473, 998
\bibitem[]{} 
     In'T Zand, J. J. M. and Fenimore, E. 1996, ApJ, 464, 662
\bibitem[]{} 
     Kim, C., Chang, H.-Y., and Yi, I. 2001, ApJ, 548, 532
\bibitem[]{} 
     Klebesadel, R. W., Strong, I. B., and Olson, R. A., 1973, ApJ 182, L85
\bibitem[]{} 
     Kobayashi, S., Piran, T., and Sari, R. 1997, ApJ, 490, 92
\bibitem[]{} 
     Lamb, D. Q. and Reichart, D. E. 2000, ApJ, 536, 1
\bibitem[]{} 
     MacFadyen, A. I. and Woosley, S. E., 1999, ApJ, 524, 262
\bibitem[]{} 
     Madau, P., et al. 1996, MNRAS, 283, 1388
\bibitem[]{}
     Mao, S. and Paczy${\rm \acute{n}}$ski, B.  1992, ApJ, 388, L45
\bibitem[]{} 
     Meegan, C. A., et al.  1992, Nature, 355, 143
\bibitem[]{} 
     Metzger, M. R., et al.  1997, Nature, 387, 879
\bibitem[]{} 
     Norris, J. P., et al. 1996, ApJ, 459, 393
\bibitem[]{} 
     Norris, J. P., Marani, G. F., and Bonnell, J. T. 2000, ApJ, 534, 248
\bibitem[]{} 
     Ostriker, J. P. and Gnedin, N. Y. 1996, ApJ, 472, L63
\bibitem[]{}
     Paciesas, W. S., et al. 1999, ApJS, 122, 465
\bibitem[]{} 
     Paczy${\rm \acute{n}}$ski, B. 1998, ApJ, 494, L45
\bibitem[]{} 
     Panaitescu, A., Spada, M., and ${\rm M\acute{e}sz\acute{a}ros }$, 
     P. 1999, ApJ, 522, L105
\bibitem[]{}
     Piran, T. 1992, ApJ, 389, L45
\bibitem[]{} 
     Reichart, D. E., et al. 2001, ApJ, 552, 57
\bibitem[]{} 
     Rowan-Robinson, M. 1999, Ap\&SS, 266, 291
\bibitem[]{} 
     Steidel, C. C., et al. 1999, ApJ, 519, 1
\bibitem[]{} 
     Stern, B., Poutanen, J., and Svensson, R.  1999, ApJ, 510, 312
\bibitem[]{}
     Wijers, R. A. M. J., et al.  1998, MNRAS, 294, L13
\bibitem[]{}
     Woosley, S. E. 1993, ApJ, 405, 273
\end{thebibliography}
\end{document}